\documentclass[preprint,prd,amsmath,amssymb,aps,nofootinbib]{revtex4-1}

\usepackage{graphicx}
\usepackage{siunitx}
\usepackage{bm}
\usepackage{natbib}
\graphicspath{{figures/}}

\def\aap{Astron.\ Astrophys.\ }
\def\apj{Astrophys.\ J.\ }

\def\mnras{Mon.\ Not.\ R.\ Astron.\ Soc.\ }

\def\prl{Phys.\ Rev.\ Lett.\ }

\begin{document}
\title{Effect of magnetic field correlation length on the gamma-ray pulsar halo morphology under anisotropic diffusion}
\author{Kun Fang$^{a}$}
\author{Hong-Bo Hu$^{a,b}$}
\author{Xiao-Jun Bi$^{a,b}$}
\author{En-Sheng Chen$^{a,b}$}

\affiliation{
$^a$Key Laboratory of Particle Astrophysics, Institute of High Energy
Physics, Chinese Academy of Sciences, Beijing 100049, China \\
$^b$University of Chinese Academy of Sciences, Beijing 100049, China\\
}

\date{\today}

\begin{abstract}
Anisotropic diffusion is one of the potential interpretations for the morphology of the Geminga pulsar halo. It interprets the observed slow-diffusion phenomenon through a geometric effect, assuming the mean magnetic field direction around Geminga is closely aligned with the line of sight toward it. However, this direction should not extend further than the correlation length of the turbulent magnetic field $L_c$, which could be $100$~pc or less. We first revisit the $L_c=\infty$ scenario and show that the halo asymmetry predicted by this scenario is mainly contributed by the electrons located beyond the ``core" section around Geminga, which has a length of $100$~pc. Then, considering the directional variation of the magnetic field beyond the core section, we take one magnetic field configuration as an example to investigate the possible halo morphology. The predicted morphology has some different features compared to the $L_c=\infty$ scenario. The current experiments may already be able to test these features. In addition, we use a semi-analytical method to solve the anisotropic propagation equation, which offers significant convenience compared to numerical approaches.
\end{abstract}

\maketitle

\section{Introduction}
\label{sec:intro}
Pulsar halos are inverse Compton (IC) gamma-ray sources generated by high-energy electrons and positrons\footnote{\textit{Electrons} will denote both electrons and positrons hereafter if not specified.} escaping from pulsar wind nebulae (PWNe) and diffusing in the interstellar medium (ISM) around the pulsars (see Ref.~\cite{Fang:2022fof,Liu:2022hqf,Lopez-Coto:2022igd} for reviews). The most notable feature of pulsar halos is the extremely slow electron diffusion rate inside the halo regions \cite{Abeysekara:2017old,Aharonian:2021jtz,Fang:2022qaf,HAWC:2023jsq}, which also enables their visibility. Pulsar halos are ideal probes for studying cosmic ray propagation in localized Galactic regions. They are believed to have critical implications on the problems of the cosmic positron excess and the diffuse TeV gamma-ray excess \cite{Hooper:2017gtd,Fang:2018qco,2018PhRvL.120l1101L}, assuming we properly understand the slow-diffusion phenomenon. 

As the diffusion coefficient is inversely proportional to the energy density of the turbulent magnetic field \cite{1971ApJ...170..265S}, the slow-diffusion phenomenon may correspond to a significant turbulence injection. The escaping electron-positron pairs themselves can excite turbulent waves through streaming instability \cite{Evoli:2018aza,Mukhopadhyay:2021dyh}, while the injection power may be too weak to significantly suppress the diffusion coefficient considering the pulsar motion \cite{Kun:2019sks}. The parent supernova remnants (SNRs) of the pulsars are promising sources of the required turbulence \cite{Kun:2019sks,Mukhopadhyay:2021dyh}, although among the known pulsar halos, only the Monogem halo has an observable associated SNR \cite{1996ApJ...463..224P,2018MNRAS.477.4414K}. The slow diffusion inside the halo regions is inferred from the steep gamma-ray profiles of the halos. It is also suggested that the steep profiles may be reproduced with a typical diffusion rate in the Galaxy if a relativistic correction to the diffusion equation is considered \cite{Recchia:2021kty}. However, the needed electron energy, in this case, is larger than what the pulsars could provide, and the goodness of fit to the data is significantly poorer than the slow-diffusion model \cite{Bao:2021hey}.

Another interpretation that does not require a strong turbulent environment is the anisotropic diffusion model \cite{Liu:2019zyj,DeLaTorreLuque:2022chz}, where the diffusion coefficient perpendicular to the mean magnetic field can be much smaller than parallel. If the mean magnetic field around a pulsar coincides with the line of sight (LOS), perpendicular diffusion could explain the steep profile of the halos, while parallel diffusion remains typical for the Galaxy. For the Geminga halo, the canonical pulsar halo, the large-scale Galactic field is not aligned with the LOS toward it. However, the turbulent field direction could significantly deviate from the large-scale field at the turbulence correlation length $L_c$ ($\lesssim100$~pc), and alignment is still possible considering this fluctuation \cite{Liu:2019zyj}.

Along this line of thought, the mean-field direction beyond $\sim L_c$ from the pulsar should no longer align with our LOS. In other words, the finiteness of $L_c$ must be considered when calculating the pulsar halo morphology, which was not specifically discussed in previous works \cite{Liu:2019zyj,DeLaTorreLuque:2022chz}. In this work, we take the Geminga halo as the object to study the impact of a finite $L_c$ on the pulsar halo morphology under the anisotropic diffusion assumption. In Sec.~\ref{sec:solution}, we describe our calculation and introduce a semi-analytical method for solving the anisotropic diffusion equation, which is much more convenient than the numerical method adopted by the previous works \cite{Liu:2019zyj,DeLaTorreLuque:2022chz}. In Sec.~\ref{sec:old_model} and \ref{sec:asymmetry}, we revisit the model that does not consider the finiteness of $L_c$ and show that the halo asymmetry expected by this scenario is actually contributed by the electrons located outside the typical scale of $L_c$ around the pulsar. As the direction of the mean magnetic field far away from Geminga is unknown, we take one magnetic field configuration as an example to discuss the possible halo morphology in Sec.~\ref{sec:new_model}. Finally, we conclude in Sec.~\ref{sec:conclu}.

\section{Semi-analytical solution of the anisotropic diffusion equation}
\label{sec:solution}
After the accelerated electrons escape from the PWN, their propagation in the interstellar medium (ISM) can be described by the diffusion-loss equation. We solve the propagation equation to obtain the electron number density around the pulsar and then do the LOS integration to obtain the electron surface density, with which the gamma-ray emission can be derived from the standard IC scattering calculation \cite{Blumenthal:1970gc}.

The electron propagation equation under anisotropic diffusion can be expressed by
\begin{equation}
 \begin{aligned}
  \frac{\partial N(E_e, r, z, t)}{\partial t} = &  \frac{D_{rr}(E_e)}{r}\frac{\partial}{\partial r}\left[r\frac{\partial N(E_e, r, z, t)}{\partial r}\right] + D_{zz}(E_e)\frac{\partial^2 N(E_e, r, z, t)}{\partial z^2} \\
  & + \frac{\partial [b(E_e)N(E_e, r, z, t)]}{\partial E_e} + Q(E_e, r, z, t) \,,
 \end{aligned}
 \label{eq:prop}
\end{equation}
where $N$ is the electron number density, $E_e$ is the electron energy, $r$ and $z$ are the coordinates perpendicular and parallel to the mean magnetic field with the pulsar position as the origin, and $t$ is the time coordinate with the pulsar birth time as the origin. As suggested by Ref.~\cite{Yan:2007uc} and \cite{Xu:2013ppa}, the diffusion coefficient perpendicular and parallel to the mean magnetic field have the relation of $D_{rr}=D_{zz}M_A^4$, where $M_A$ is the Alfv\'{e}nic Mach number. The perpendicular diffusion coefficient is smaller than the parallel one in the sub-Alfv\'{e}nic regime. The energy-loss rate is denoted with $b(E_e)=b_0(E_e)E_e^2$. We take a magnetic field strength of 3~$\mu$G for the synchrotron loss rate. We adopt the background photon fields given in Ref.~\cite{Abeysekara:2017old} and the parametrization method given in Ref.~\cite{Fang:2020dmi} to calculate the IC loss rate.

We divide the source function $Q$ into a spatial term $q_x$, a temporal term $q_t$, and an energy term $q_E$. As the bow-shock PWN size is much smaller than the gamma-ray halo, we can safely assume a point-like source as 
\begin{equation}
 q_x(r, z)=\frac{1}{r}\delta(r)\delta(z)\,.
 \label{eq:spatial}
\end{equation}
The temporal term is assumed to follow the variation of the pulsar spin-down luminosity as
\begin{equation}
 q_t(t)=\left\{
 \begin{aligned}
  & [(1+t/t_{\rm sd})/(1+t_p/t_{\rm sd})]^{-2}\,, \quad & t\geq0 \\
  & 0\,, \quad & t<0 \\
 \end{aligned}
 \right.\,,
 \label{eq:temporal}
\end{equation}
where $t_p$ is the pulsar age, and $t_{\rm sd}$ is the pulsar spin-down time scale set to be $10$~kyr. The injection energy spectrum is assumed to be a power law with an exponential cutoff as
\begin{equation}
 q_E(E_e)=q_{e,0}\,E_e^{-p}\,{\rm exp}\left[-\left(\frac{E_e}{E_{e,c}}\right)^2\right]\,,
 \label{eq:inj}
\end{equation}
which is suggested by the relativistic shock acceleration theory of electrons \cite{Dempsey:2007ng}.

We rescale the $r$ coordinate to $r'=ar$, where $a=M_A^{-2}$. Then Eq.~(\ref{eq:prop}) can be rewritten as
\begin{equation}
 \begin{aligned}
  \frac{\partial N'(E_e, r', z, t)}{\partial t} = &  \frac{D_{zz}(E_e)}{r'}\frac{\partial}{\partial r'}\left[r'\frac{\partial N'(E_e, r', z, t)}{\partial r'}\right] + D_{zz}(E_e)\frac{\partial^2 N'(E_e, r', z, t)}{\partial z^2} \\
  & + \frac{\partial [b(E_e)N'(E_e, r', z, t)]}{\partial E_e} + Q'(E_e, r', z, t) \,,
 \end{aligned}
 \label{eq:prop2}
\end{equation}
where $N'(E_e, r', z, t)=N(E_e, r'/a, z, t)$, and $Q'(E_e, r', z, t)=q_x'(r', z)q_E(E_e)q_t(t)$, where
\begin{equation}
 q'_x(r', z)=q_x(r'/a, z)=\frac{a^2}{r'}\delta(r')\delta(z)\,.
 \label{eq:spatial2}
\end{equation}
Now Eq.~(\ref{eq:prop2}) describes \textit{isotropic} diffusion in cylindrical coordinates, and we can straightforwardly write the solution given by the image method \cite{2010A&A...524A..51D} as
\begin{equation}
 \begin{aligned}
  N'(E_e, r', z, t) = & \int_{t_{\rm ini}}^tdt_0\iint r'_0\,dr'_0\,dz_0\,Q'(E_e^\star, r'_0, z_0, t_0)\,\frac{b(E_e^\star)}{b(E_e)}\,\frac{1}{(\pi\lambda^2)^{3/2}} \\ 
  & \times {\rm exp}\left[-\frac{(r'-r'_0)^2}{\lambda^2}\right]\,\sum_{n=-\infty}^{+\infty}{\rm exp}\left\{\frac{[z-(-1)^nz_0-2nz_{\rm max}]^2}{\lambda^2}\right\} \\
  = & \int_{t_{\rm ini}}^tdt_0\,q_t(t_0)\,q_E(E_e^\star)\,\frac{b(E_e^\star)}{b(E_e)}\,\frac{a^2}{(\pi\lambda^2)^{3/2}}\,{\rm exp}\left(-\frac{r'^2}{\lambda^2}\right)\,\sum_{n=-\infty}^{+\infty}{\rm exp}\left[\frac{(z-2nz_{\rm max})^2}{\lambda^2}\right]\,, \\
 \end{aligned}
 \label{eq:solution}
\end{equation}
where 
\begin{equation}
 E_e^\star\approx\frac{E_e}{[1-b_0E_e(t-t_0)]}\,,\quad\lambda^2=4\int_{E_e}^{E_e^\star}\frac{D_{zz}(E'_e)}{b(E'_e)}\,dE'_e\,,\quad t_{\rm ini}={\rm max}\{t-1/(b_0E_e), 0\}\,,
 \label{eq:definite}
\end{equation}
and $z_{\rm max}$ is set to be $2$~kpc, which is significantly larger than $\lambda$ within the energy range of interest. Finally, we obtain the solution of Eq.~(\ref{eq:prop}), which is written as
\begin{equation}
  N(E_e, r, z, t) = \int_{t_{\rm ini}}^tdt_0\,q_t(t_0)\,q_E(E_e^\star)\,\frac{b(E_e^\star)}{b(E_e)}\,\frac{a^2}{(\pi\lambda^2)^{3/2}}\,{\rm exp}\left(-\frac{a^2r^2}{\lambda^2}\right)\,\sum_{n=-\infty}^{+\infty}{\rm exp}\left[\frac{(z-2nz_{\rm max})^2}{\lambda^2}\right]\,.
 \label{eq:solution2}
\end{equation}

We denote the angle between the z axis and the direction to the pulsar with $\Phi$. For the LOS direction with an angle of $\theta$ away from the pulsar, the electron surface density can be calculated by
\begin{equation}
 S_e(\theta)=\int N(r(l, \theta, \Phi), z(l, \theta, \Phi))dl\,.
\end{equation}
One may refer to Ref.~\cite{Liu:2019zyj} for the transformation between $(l, \theta, \Phi)$ and $(r, z)$.

If the finiteness of $L_c$ is considered, the $z$ axis should change direction significantly at each $L_c$. We take an equivalent approach to this problem in the LOS integration step, which will be described in detail in Sec.~\ref{sec:new_model}.

\section{Anisotropic diffusion model with infinite magnetic field correlation length}
\label{sec:old_model}
\begin{figure}[t!]
\begin{center}
\includegraphics[width=12cm]{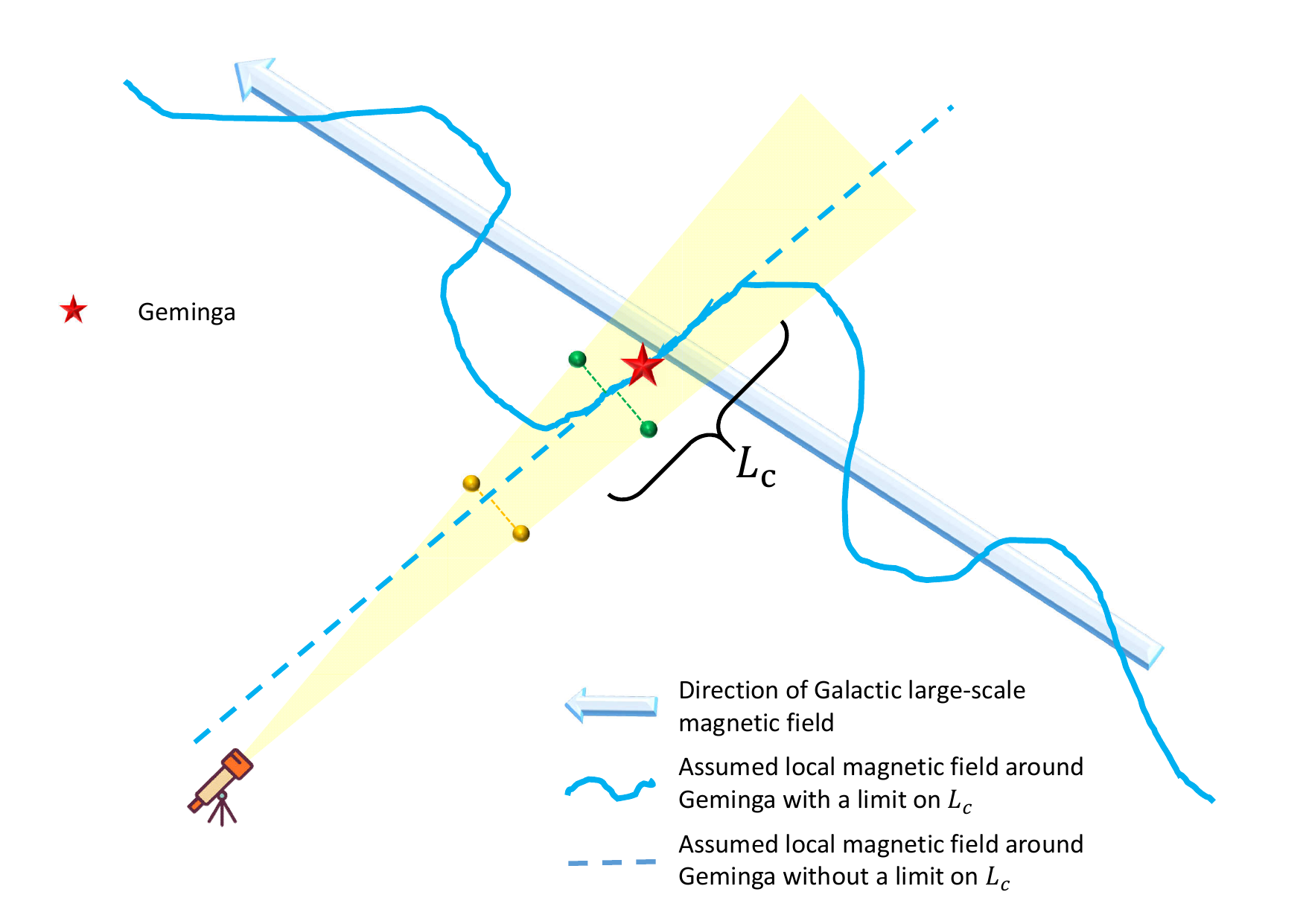}
\end{center}
\caption{Schematic diagram showing a possible relationship between the directions of the Galactic large-scale magnetic field, the turbulent magnetic field around Geminga, and our LOS toward Geminga. The correlation length of the turbulent field $L_c$ is assume to be $\sim100$~pc, significantly smaller than the distance between the observer and Geminga.}
\label{fig:sketch}
\end{figure}

Figure~\ref{fig:sketch} shows a possible magnetic field configuration around Geminga for the anisotropic-diffusion interpretation. The Galactic large-scale magnetic field near the Galactic plane should follow the direction of the local spiral arm (the Orion Spur), which deviates significantly from the LOS toward Geminga. However, considering the fluctuation of the turbulent field at the scale of $L_c$, there is a possibility that the mean field around Geminga coincides with our LOS, although the field should change direction significantly outside the scale of $L_c$ (the solid blue line in Fig.~\ref{fig:sketch}).

\begin{figure}[t!]
\begin{center}
\includegraphics[width=8cm]{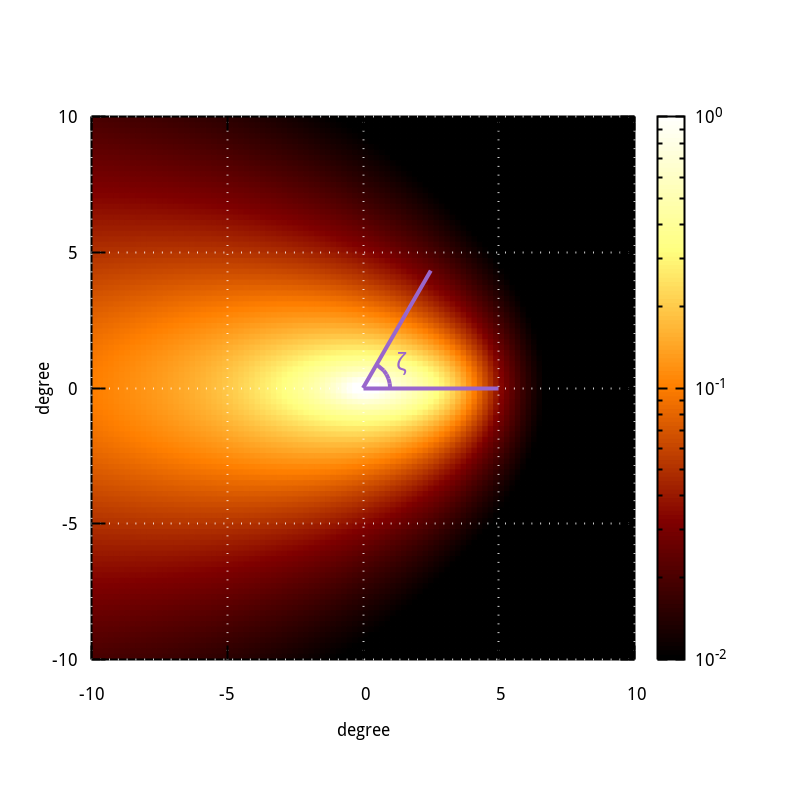}
\includegraphics[width=8cm]{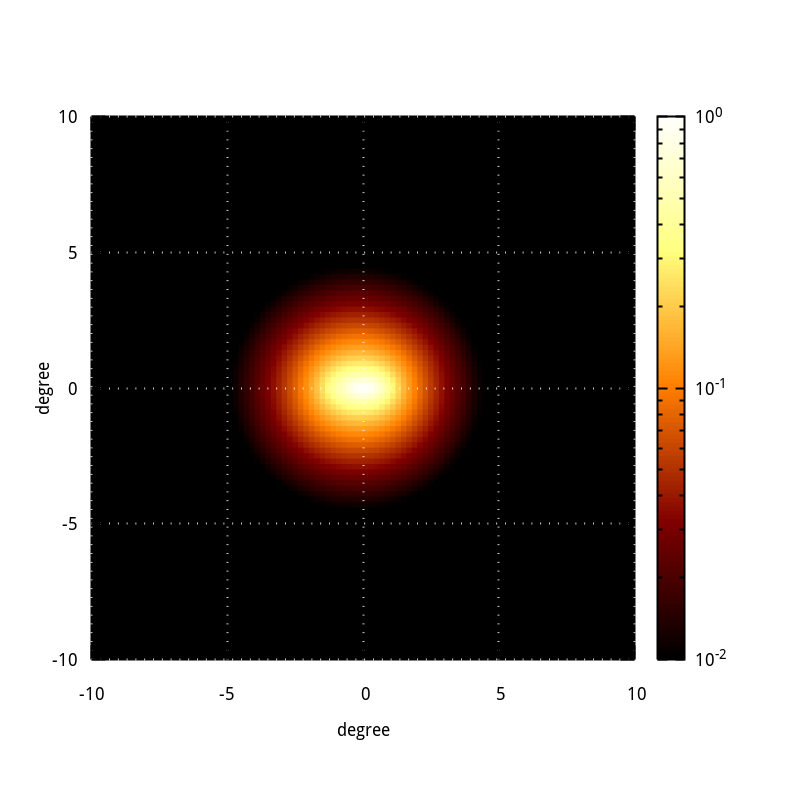}
\end{center}
\caption{Left: Gamma-ray morphology of the Geminga halo expected by the anisotropic diffusion model, assuming $L_c=\infty$, $\Phi=5^\circ$, $M_A=0.2$, and $D_{rr,100}=3\times10^{27}$~cm$^2$~s$^{-1}$. The units of the color bar are $10^{-12}$~TeV~cm$^{-2}$~s$^{-1}$~deg$^{-2}$. The energy integration range is $8-40$~TeV. Right: Same as the left, but only consider the contribution of the electrons located within the ``core" section depicted in Fig.~\ref{fig:sketch2}.}
\label{fig:map}
\end{figure}

In the previous calculations of the anisotropic diffusion model \cite{Liu:2019zyj,DeLaTorreLuque:2022chz}, the magnetic field in the ISM always remains in its direction near Geminga, as shown by the dashed blue line in Fig.~\ref{fig:sketch}. It means that the finiteness of $L_c$ is not considered or that $L_c$ is assumed to be much larger than indicated by observations \cite{Haverkorn:2008tb,Iacobelli:2013fqa}. In this section, we revisit this scenario by comparing it with the HAWC measurement of the Geminga halo \cite{Abeysekara:2017old}. The HAWC collaboration did not report significant asymmetry in the Geminga halo, while the most distinct feature of anisotropic diffusion is its expected asymmetry. The left panel of Fig.~\ref{fig:map} shows an example of the Geminga halo morphology with $\Phi=5^\circ$. The asymmetry increases as $\Phi$ increases. 

For a specific $\Phi$, we test if the asymmetry could be identified with the data size used in the original paper of HAWC \cite{Abeysekara:2017old}. In the initial step, we calculate the average gamma-ray profile in $0^\circ<\zeta<360^\circ$ to fit the surface brightness profile given by HAWC, where $\zeta$ is the azimuth marked in the left panel of Fig.~\ref{fig:map}. The main parameters of the model are determined by this fitting procedure. Then in the subsequent step, we calculate the integrated fluxes in three different intervals of $0^\circ<\zeta<60^\circ$, $60^\circ<\zeta<120^\circ$, and $120^\circ<\zeta<180^\circ$ and test if there is a significant difference between them. 

The diffusion coefficient take the form of $D_{i}=D_{i,100}(E/{\rm 100~TeV})^\delta$, where $i=zz$ or $rr$. The slope is assumed to be $\delta=1/3$, as suggested by Kolmogorov's theory. The parallel diffusion coefficient $D_{zz}$ should be consistent with the cosmic-ray boron-to-carbon ratio (B/C) measurements \cite{Aguilar:2016vqr,Collaboration:2022vwu,CALET:2022dta}. The DAMPE experiment measures the B/C up to $\approx5$~TeV/n and finds a spectral hardening at $\approx100$~GeV/n. Assuming the B/C spectrum above the hardening can extrapolate to 100~TeV, and the spectral hardening is entirely attributed to the slope change of the diffusion coefficient, we can get a lower limit of $D_{zz,100}$ of $\approx7\times10^{29}$~cm$^2$~s$^{-1}$ (see model B' of Ref.~\cite{Ma:2022iji}). We set $D_{zz,100}$ as a free parameter in the fitting procedure. Another free parameter is $M_A$, which together with $D_{zz,100}$ determines $D_{rr,100}$. 

For the injection spectrum, we set $p=1.0$ and $E_{e,c}=130$~TeV as suggested by a fit to the HAWC gamma-ray spectrum \cite{Bao:2021hey}. The conversion efficiency from the pulsar spin-down energy to the injected electron energy is set to be a free parameter and denoted with $\eta$, which mainly determines the normalization of the injection spectrum. Since the gamma-ray profile provided by HAWC is in a single energy bin of $8-40$~TeV, fixing the parameters that mainly determine energy-dependent features ($p$, $E_{e,c}$, and $\delta$) should hardly impact our results.

\begin{figure}[t!]
\begin{center}
\includegraphics[width=8cm]{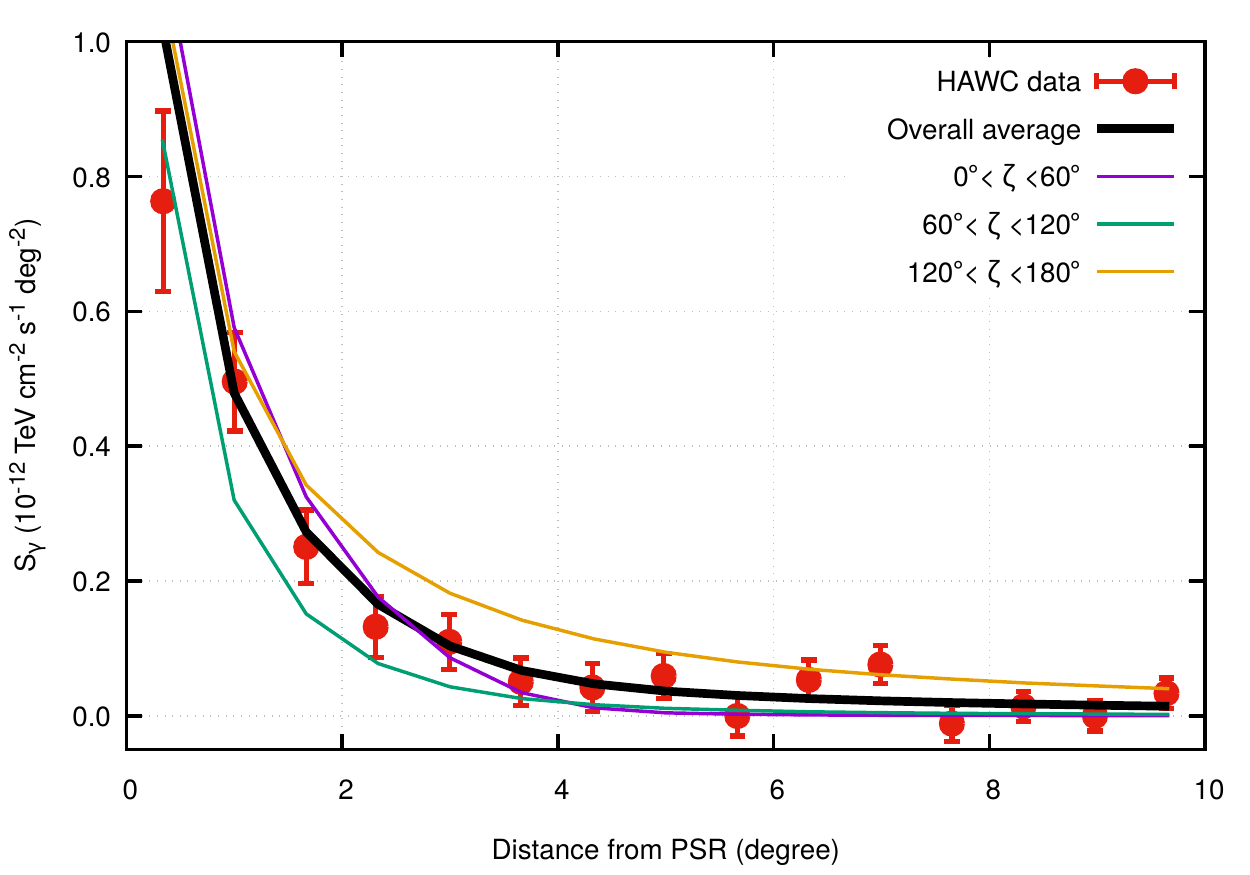}
\includegraphics[width=8cm]{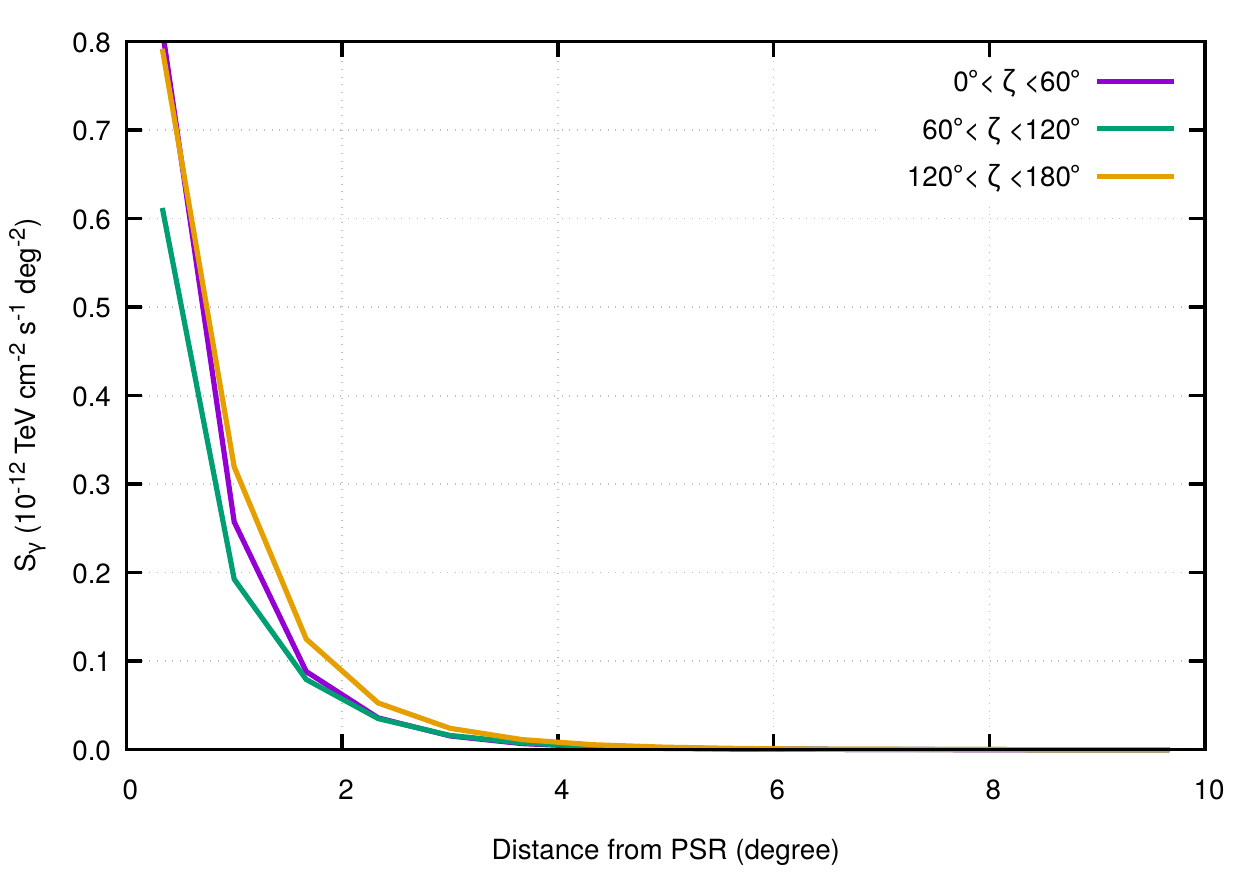}
\end{center}
\caption{Left: Fitting result to the HAWC measurement of the Geminga halo profile \cite{Abeysekara:2017old} assuming $\Phi=5^\circ$ and $L_c=\infty$. Profiles in different azimuth intervals calculated with the best-fit parameters are also shown. Right: Based on the subinterval profiles in the left panel, but only consider the contribution of the electrons located within the ``core" section depicted in Fig.~\ref{fig:sketch2}.}
\label{fig:prof}
\end{figure}

As an example, we show the fitting result for $\Phi=5^\circ$ in the left panel of Fig.~\ref{fig:prof}. The best-fit parameters are $D_{zz,100}=1.8\times10^{30}$~cm$^2$~s$^{-1}$, $M_A=0.17$, and $\eta=8.2\%$. It can be seen that the full-azimuth averaged profile can fit the HAWC data well (reduced $\chi^2$ is $0.89$). However, the expected profiles in different subintervals are clearly different.

To estimate the significance of predicted flux differences between the subintervals, we need to determine the statistical errors in the measurements. We convert the HAWC measured flux into event counts and estimate the background. The live time is set to be 507 days \cite{Abeysekara:2017old}, and the observation time of the Geminga halo is about 6 hours on each transit. The average photon energy of the measurement is 20~TeV \cite{Abeysekara:2017old}. According to of HAWC effective area of $22000$~m$^2$ \cite{Zhou:2021dgj}, the mean flux value can be converted to $n_s$, and the flux error roughly to $\sqrt{n_s\Omega+n_b\Omega}/\Omega$, where $n_s$ and $n_b$ are the number of excess events and background events per square degree, respectively, and $\Omega$ is the solid angle of an angular interval. We assume that $n_b$ is uniform in all the angular intervals and obtain an average $n_b$ of $405$~deg$^{-2}$.

\begin{figure}[t!]
\begin{center}
\includegraphics[width=10cm]{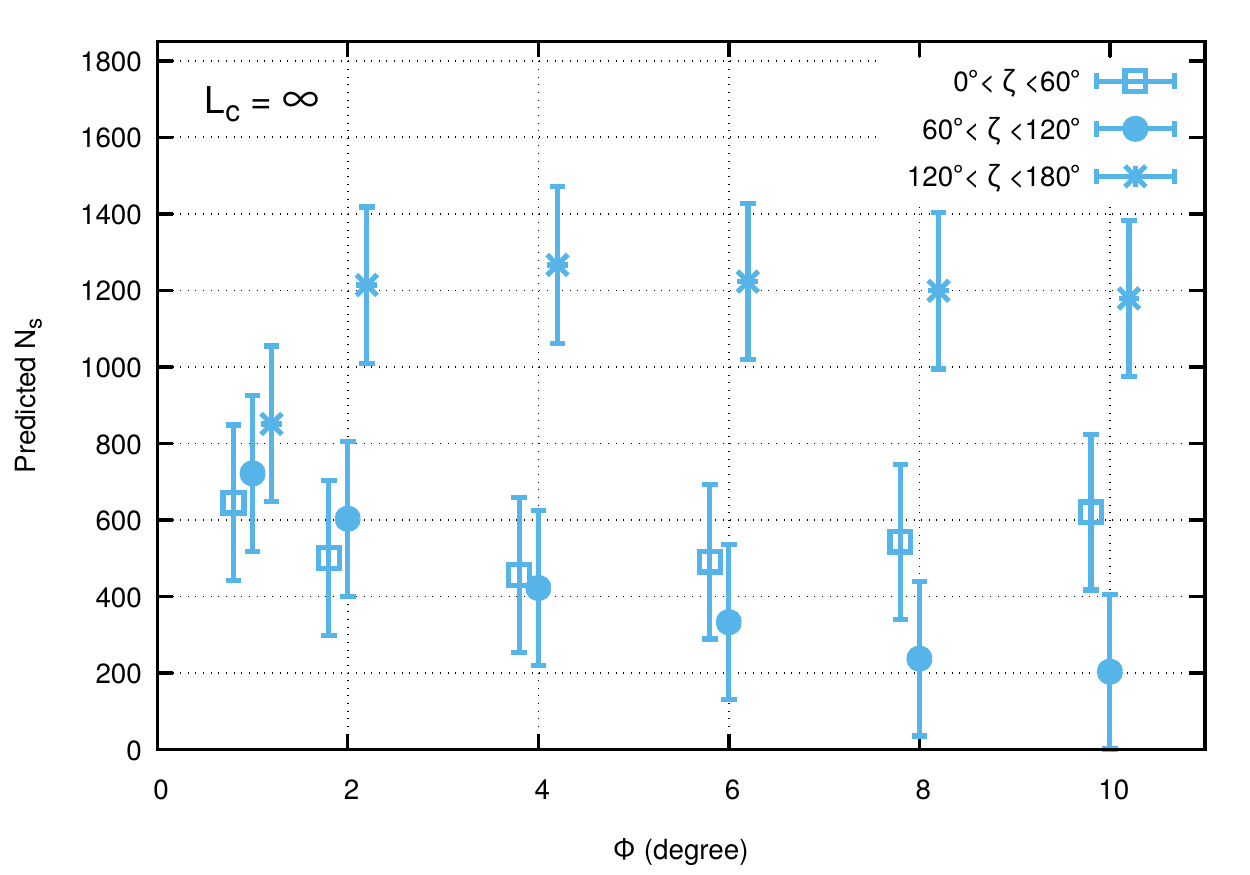}
\end{center}
\caption{Number of excess events in different regions of the Geminga halo predicted by the anisotropic diffusion model, assuming an infinite extent for the mean magnetic field around Geminga ($L_c=\infty$). The experiment parameters adopted in the HAWC paper \cite{Abeysekara:2017old} is used for the estimation.}
\label{fig:flux_test}
\end{figure}

For each subinterval, the predicted excess integrated within $10^\circ$ around the pulsar can be expressed as $N_s\approx N_{\rm model}\pm\sqrt{N_{\rm model}+N_b}$, where $N_{\rm model}$ is the integrated model value, and $N_b$ is the integrated background. We compare the predicted $N_s$ in different subintervals\footnote{As the model is symmetrical relative to the horizontal axis (see Fig.~\ref{fig:map}), the region of $0^\circ<\zeta<60^\circ$ actually represents $-60^\circ<\zeta<60^\circ$. The same is true for other subintervals.} in Fig.~\ref{fig:flux_test}. As shown, the difference between the subintervals of $0^\circ<\zeta<60^\circ$ and $120^\circ<\zeta<180^\circ$ could be identified with a significance of $3\sigma$ for $\Phi=2^\circ$. The asymmetry is more significant for larger $\Phi$ as expected. For $\Phi=1^\circ$, the excess difference is no longer significant between the subintervals, which may not be detectable with the HAWC data. This restriction given by this integrated-flux test could be more stringent than Ref.~\cite{DeLaTorreLuque:2022chz}, where $\Phi$ is constrained to be smaller than $\approx2.5^\circ$.

\section{Origin of the predicted halo asymmetry}
\label{sec:asymmetry}
Observations indicate that the correlation length of the turbulent magnetic field in the Galactic ISM falls in $1-100$~pc \cite{Haverkorn:2008tb,Iacobelli:2013fqa}. The correlation length is determined by the turbulence injection scale, which is several times smaller than the actual injection scale \cite{Harari:2002dy}. As the main turbulence sources, SNRs typically have scales of several dozen parsecs, so it is unlikely that $L_c$ near Geminga is significantly larger than $100$~pc, given that no large-scale structure is found in that region. Furthermore, the mean magnetic field within $20$~pc around the solar system, as measured by the Interstellar Boundary Explorer (IBEX), forms a $\approx60^\circ$ angle with the direction of Geminga \cite{2013ApJ...776...30F}. This local field direction could be distorted by the Local Bubble \cite{2018A&A...611L...5A}. Thus, even if the mean field near Geminga coincides with our LOS, it is unlikely that it extends all the way to the vicinity of the solar system.

\begin{figure}[t!]
\begin{center}
\includegraphics[width=10cm]{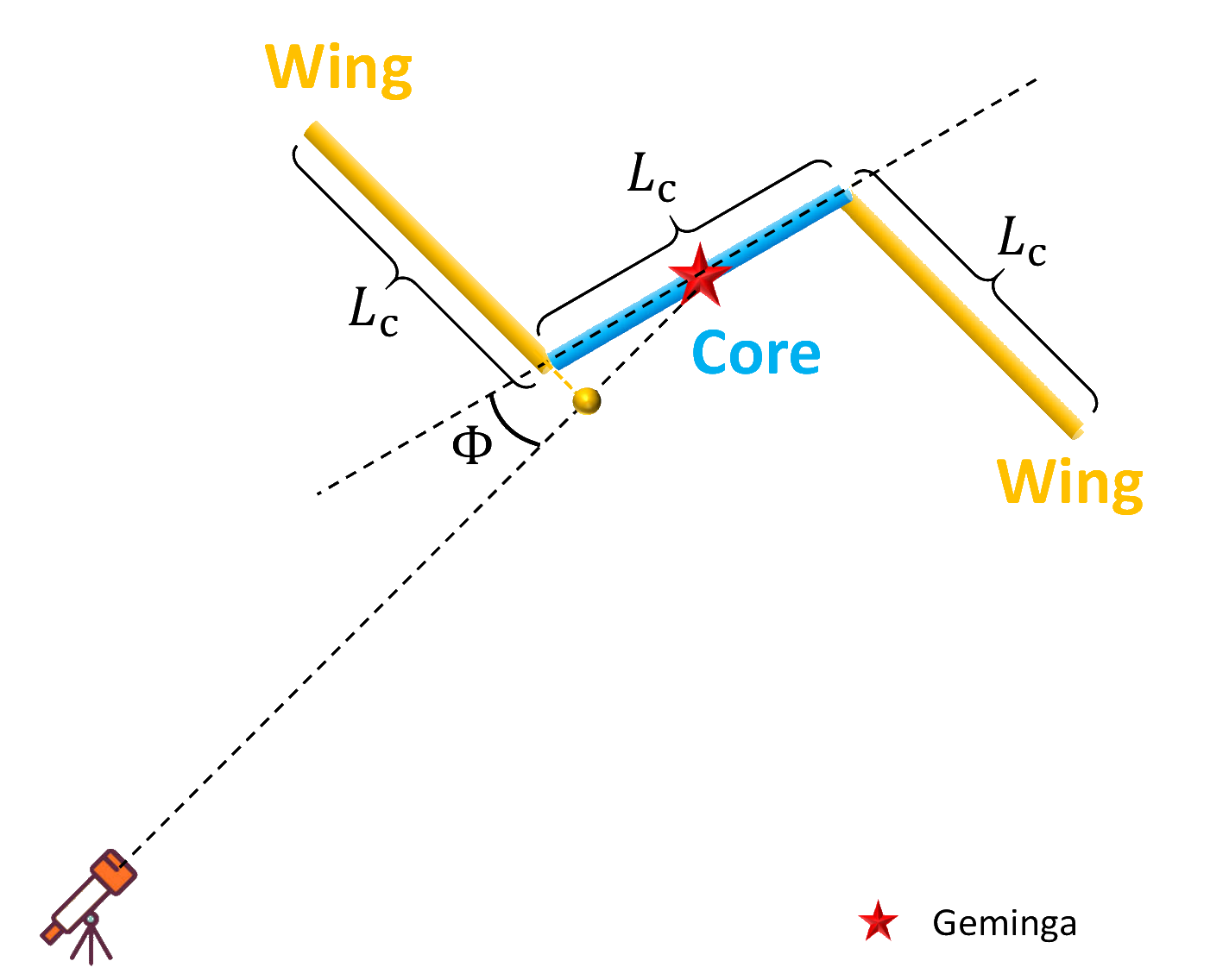}
\end{center}
\caption{Schematic diagram showing a possible magnetic field configuration in the ISM around Geminga. Electrons propagating along the core part, which roughly aligns with the LOS and has a length of $L_c=100$~pc, may interpret the observed steep profile of the Geminga halo. Electrons leaving the core part will be deflected to the directions of the wing parts, contributing additional asymmetry to the halo morphology.}
\label{fig:sketch2}
\end{figure}

To interpret the halo morphology with the anisotropic diffusion model, the mean magnetic field direction within the ``core" section around Geminga, which has a length of $L_c$ as depicted in Fig.~\ref{fig:sketch2}, must align with the LOS direction. We assume $L_c=100$~pc in the following calculations of the paper unless specified. We evaluate the gamma-ray contribution of electrons located within the core section, which is referred to as the core component. We calculate the gamma-ray emission by only including the electrons located within the range of $-L_c/2<z<L_c/2$ in the LOS integration step. In the right panel of Fig.~\ref{fig:map}, we show an example of the gamma-ray morphology of the core component. The parameters used here are the same as those in the left panel of Fig.~\ref{fig:map}. It can be seen that the asymmetry is significantly reduced when only the core component is retained. In the right panel of Fig.~\ref{fig:prof}, we show the gamma-ray profiles in three subintervals with the same parameters used in the left panel but only include the core component in the calculation. The difference between the profiles is significantly reduced compared with the left, especially at large angles.

We provide a qualitative explanation for the above results. For the anisotropic diffusion model, the electron number density decreases slowly along the $z$ direction and rapidly along the $r$ direction. Consequently, for a position not very far from Geminga, the electron number density is mainly determined by its $r$ coordinate. We plot two sets of points that are symmetric with respect to Geminga in Fig.~\ref{fig:sketch}. The green set of points, located within the core section, exhibits a similar $r$ coordinate and hence an equivalent electron number density. This explains why the asymmetry of the core component is not significant. The yellow set of points is outside the core section. The $r$ coordinate of the point on the left is significantly smaller than that on the right, so the electron number density of the point on the left is significantly larger than that on the right. This means that the electrons outside the core section mainly determine the asymmetry feature of the halo.

Based on the analysis presented above, we highlight that a reasonable prediction of the pulsar halo morphology features requires accounting for the impact of the directional fluctuation of the mean magnetic field beyond the core section around the pulsar.

\section{Possible pulsar halo morphology considering the variation of magnetic field direction}
\label{sec:new_model}
Given that the direction of the mean magnetic field outside the core section cannot be restricted, we investigate the possible morphology of the Geminga halo under the anisotropic diffusion model based on a simple magnetic field configuration. As shown in Fig.~\ref{fig:sketch2}, the direction of the mean magnetic field experiences significant variations outside the core part, becoming perpendicular to the LOS, with symmetric variations on both sides. This implies that the propagation of electrons will undergo considerable deflection after leaving the core section. The gamma-ray emission of the electrons propagating in the direction of the yellow lines depicted in Fig.~\ref{fig:sketch2} are referred to as the ``wing" components. This scenario is referred to as the $L_c=100$~pc model.

\begin{figure}[t!]
\begin{center}
\includegraphics[width=8cm]{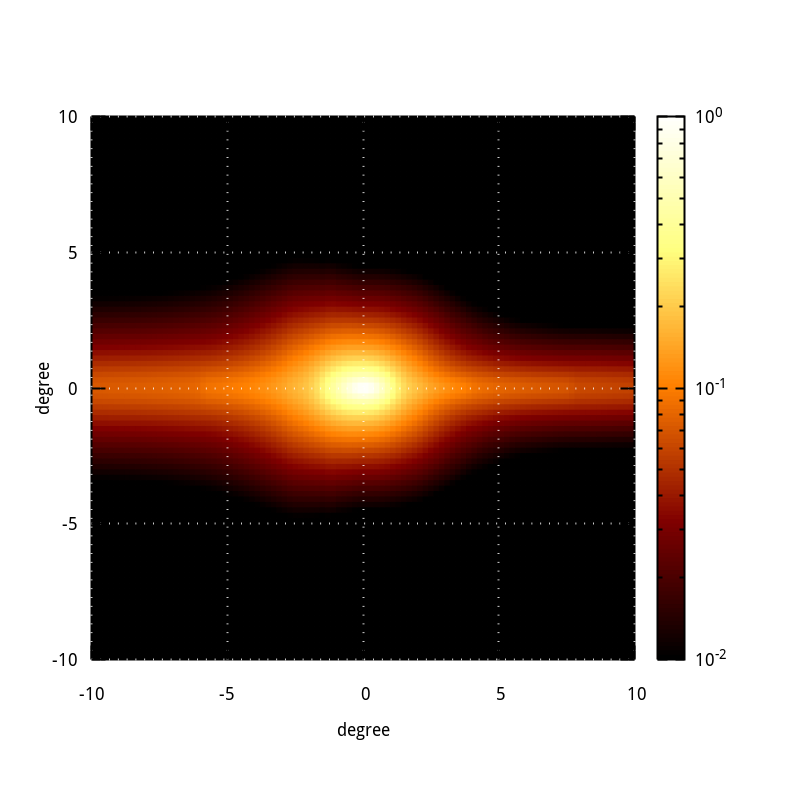}
\end{center}
\caption{Same as Fig.~\ref{fig:map}, but assuming the magnetic field configuration depicted in Fig.~\ref{fig:sketch2}.}
\label{fig:map2}
\end{figure}

The calculation of the core component has been explained in Sec.~\ref{sec:asymmetry}. For the wing components, we first assume that the $z$-axis direction remains unchanged compared to the core part and get the electron number density distribution. To achieve the effect of altering the $z$-axis direction, we transform the $z$ coordinate during the LOS integration step. For the left wing part in Fig.~\ref{fig:sketch2}, the origin of the new $z$ coordinate, $\tilde{z}$, is taken at the position of the yellow dot in Fig.~\ref{fig:sketch2}, and the $\tilde{z}$ axis is along the direction of the wing. The angle between the $\tilde{z}$ axis and the LOS is $\tilde{\Phi}=90^\circ$. The electron number density in the $\tilde{z}$ coordinate is denoted by $\tilde{N}$. Then the relationship between $\tilde{N}$ and $N$ is given by 
\begin{equation}
 \tilde{N}(\tilde{z})=N\left(z+\frac{L_c}{2}-\frac{L_c}{2}\sin\Phi\right)\,.
 \label{eq:trans}
\end{equation}
During the LOS integration of this wing component, we consider only the electron number density within the range of $(L_c/2)\sin\Phi<\tilde{z}<L_c(1+1/2\sin\Phi)$. The wing component on the right side can be obtained similarly.

In Fig.~\ref{fig:map2}, We show the morphology of the Geminga halo predicted by the above magnetic field configuration. The parameters used are the same as those in Fig.~\ref{fig:map}. It can be seen that the differential fluxes of the wing components are lower than that of the core component, so the predicted halo profile remains steep around the pulsar and can be consistent with the observation. However, due to the large angular extent of the wing components, the integrated fluxes in different azimuth intervals may exhibit notable differences. In addition, we can see that the flux of the wing component on the left is higher than that on the right due to the closer distance of the former to our observation point.

\begin{figure}[t!]
\begin{center}
\includegraphics[width=8cm]{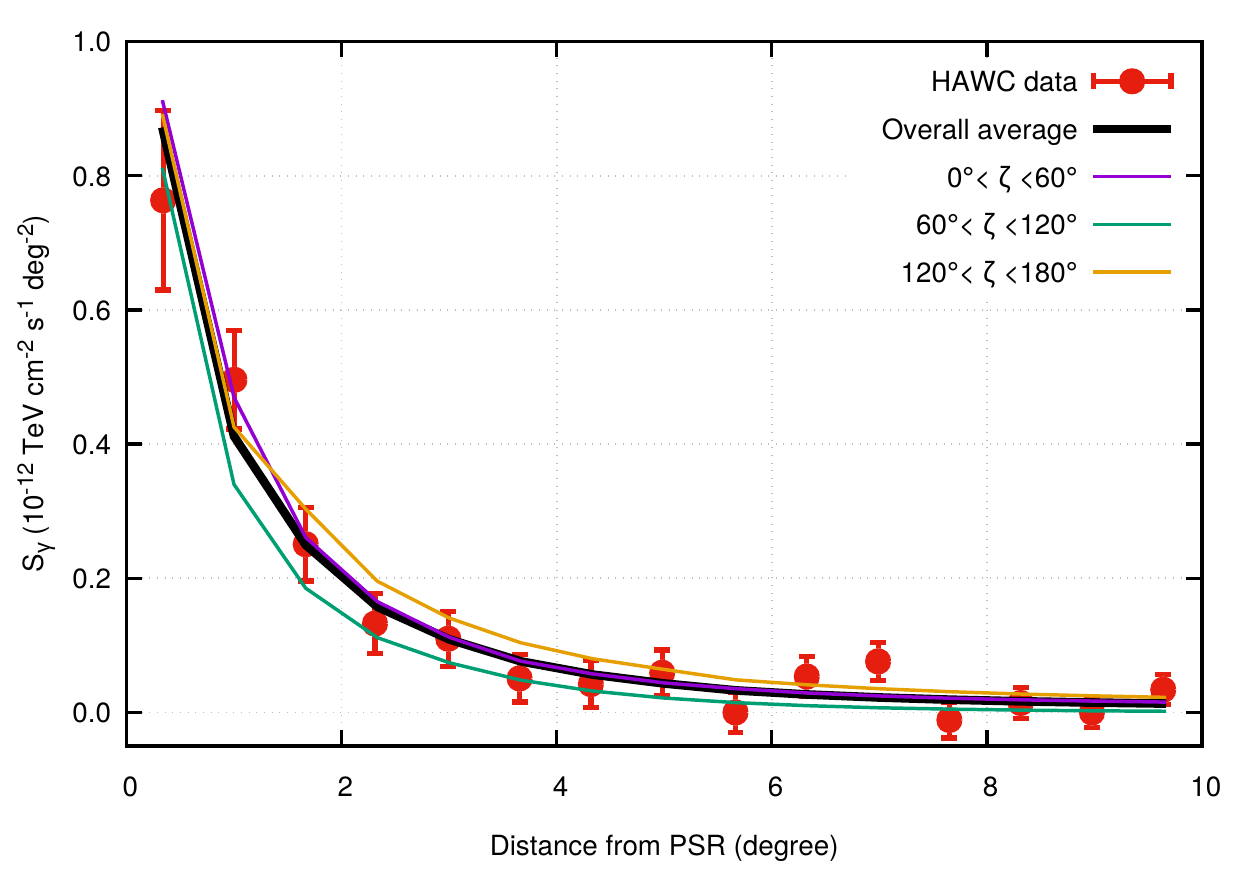}
\includegraphics[width=8cm]{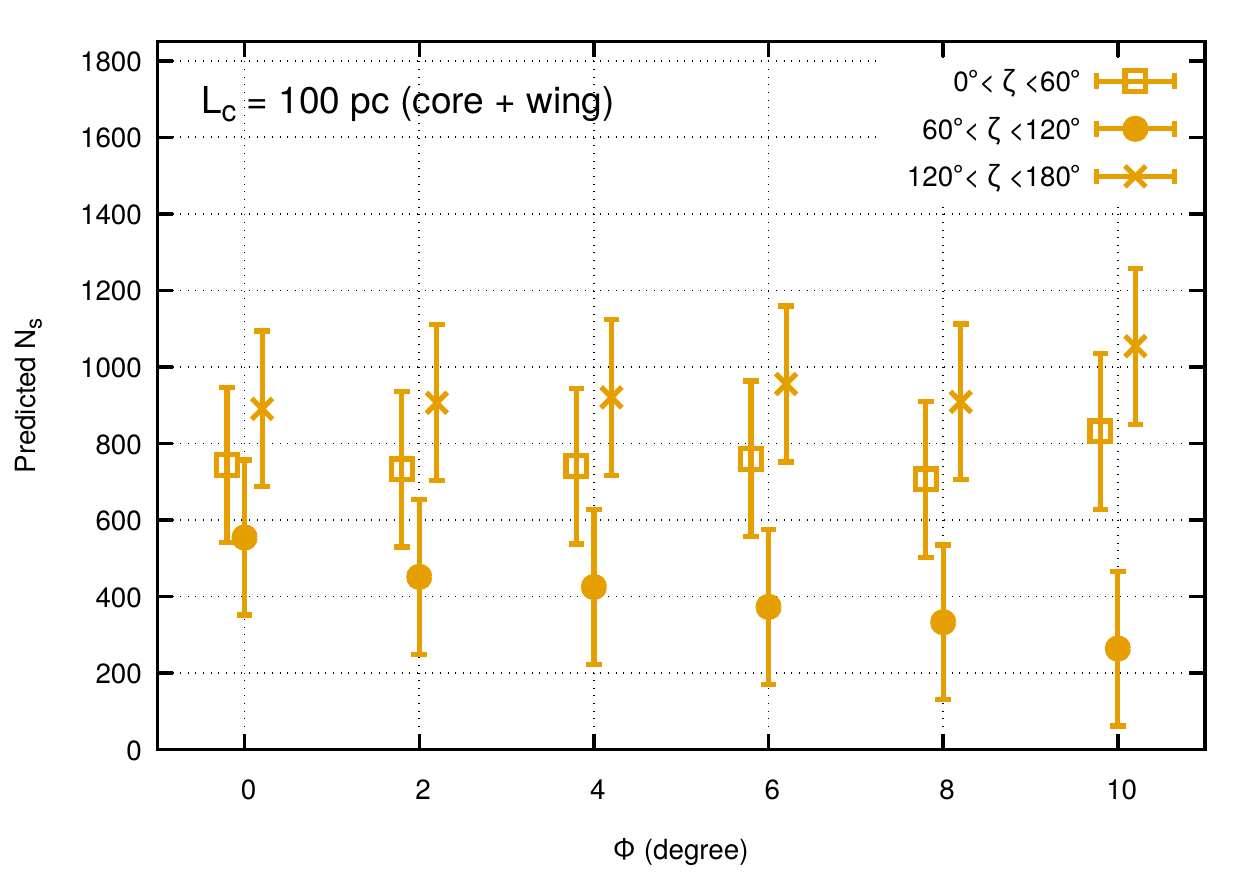}
\end{center}
\caption{Left: Same as the left panel of Fig.~\ref{fig:prof}, but assuming the magnetic field configuration depicted in Fig.~\ref{fig:map2} and $L_c=100$~pc. Right: Same as Fig.~\ref{fig:flux_test}, but assuming the magnetic field configuration depicted in Fig.~\ref{fig:map2} and $L_c=100$~pc.}
\label{fig:new_model}
\end{figure}

We repeat the calculations in Sec.~\ref{sec:old_model}, first fitting the model with various $\Phi$ to the HAWC data and then predicting the number of excess events in different azimuth intervals to determine whether the asymmetry could be detected by the integrated-flux test. Note that the direction of the wing parts is always perpendicular to the LOS, that is, $\tilde{\Phi}=90^\circ$. 

For comparison with the results presented in Fig.~\ref{fig:prof}, we show the fitting result of this new model in the left panel of Fig.~\ref{fig:new_model}, assuming $\Phi=5^\circ$. The best-fit parameters are $D_{zz,100}=1.2\times10^{30}$~cm$^2$~s$^{-1}$, $M_A=0.26$, and $\eta=19\%$. 
The difference between the predicted profiles in different subintervals is considerably smaller than that of $L_c=\infty$. As shown in the right panel of Fig.~\ref{fig:new_model}, only when $\Phi$ reaches $10^\circ$, does the difference between the predicted excesses in different subintervals has a significance of $3\sigma$.

There are several remarkable differences in the expected halo morphology between the $L_c=100$~pc and $L_c=\infty$ scenarios. First, for $\Phi$ values that are not very small, the $L_c=100$~pc model predicts a smaller halo asymmetry, which is less likely to be detected. Current observations of pulsar halos do not indicate significant asymmetry, indicating that the possibility of anisotropic diffusion explaining the pulsar halo is relatively higher after considering the finiteness of $L_c$.

Second, by comparing Fig.~\ref{fig:flux_test} and the right panel of Fig.~\ref{fig:new_model}, it is evident that for the $L_c=\infty$ scenario, the asymmetry of the halo will vanish as $\Phi$ approaches $0^\circ$, whereas for the case of $L_c=100$~pc, the halo will still retain a certain degree of asymmetry even if $\Phi=0^\circ$, due to the presence of the wing components. This implies that if the anisotropic diffusion model is correct, the asymmetry of the halo will be inevitably detected when the data size is larger. Assuming a live time of 500 days, we estimate the data size collected by the Water Cherenkov Detector Array of the Large High Altitude Air Shower Observatory (LHAASO-WCDA), which has an effective area of $\approx78000$~m$^2$ \cite{Ma:2022aau}. If $\Phi=0^\circ$, the expected number of excess events in the subinterval of $60^\circ<\zeta<120^\circ$ is $\approx1964\pm382$, and that in the subinterval of $120^\circ<\zeta<180^\circ$ is $\approx3155\pm382$. Therefore, the difference between the two regions would be detected with a significance of $3\sigma$ even for $\Phi=0^\circ$.

Third, the halo morphology expected by the $L_c=100$~pc scenario could be more complex at large angles than $L_c=\infty$. Although we show only one magnetic field configuration as an example and cannot go through all the possibilities, it is easy to infer that due to the presence of the wing components, the halo morphology will not exhibit a specific regularity as in the $L_c=\infty$ case.

If $L_c$ is several times smaller than $100$~pc, the core component will be dimmer and no longer dominate the halo morphology. The expected gamma-ray morphology will then display a strong asymmetry as illustrated by the simulations of Ref.~\cite{Lopez-Coto:2017pbk}, which may no longer explain the observations. If $L_c$ is further decreased to the order of $1$~pc, the model will regress to the slow-diffusion scenario \cite{Lopez-Coto:2017pbk} as the diffusion coefficient is positively correlated with $L_c$ when the electron Larmor radius is much smaller than $L_c$ \cite{Aloisio:2004jda}. There is no need to assume anisotropic diffusion in this case.

\section{Conclusion}
\label{sec:conclu}
In this study, we examine the anisotropic diffusion model as a potential interpretation of the pulsar halo morphology. The main point of this work is to illustrate the significance of accounting for the finiteness of $L_c$ in comparison to ignoring it, where $L_c$ is the correlation length of the turbulent magnetic field in the ISM. Our analysis focuses on the Geminga halo, a canonical example of pulsar halos.

First, we discuss the model with $L_c=\infty$, which assumes that the mean magnetic field around the pulsar is aligned closely with the LOS and extends infinitely. In the initial step, we calculate the azimuth-averaged gamma-ray profile for various $\Phi$ to fit the HAWC data. $\Phi$ denotes the angle between the mean field and the LOS. In the subsequent step, we predict the profiles in different azimuth intervals using the best-fit parameters in the previous step and then evaluate the halo asymmetry by analyzing the difference of the predicted integrated number of excess events within the subintervals, assuming the data size presented in the HAWC paper \cite{Abeysekara:2017old}. The expected asymmetry of the model would be detected with a $3\sigma$ significance as long as $\Phi\gtrsim2^\circ$.

Observations suggest that $L_c$ is on the order of $100$~pc or less. Meanwhile, we find that the expected asymmetry in the $L_c=\infty$ scenario mainly comes from the contribution of the electrons located beyond the ``core" section around the pulsar, which has a length of $100$~pc. Considering the finiteness of $L_c$, the electron propagation beyond the core part should have significantly deviated from the LOS. Therefore, we highlight that in order to reasonably predict the pulsar halo morphology in the anisotropic diffusion model, it is necessary to consider the variation of the field direction beyond the core section.

Given that the field direction beyond the core part cannot be restricted, we assume one simple magnetic field configuration and $L_c=100$~pc to investigate the possible morphology of the Geminga halo under anisotropic diffusion. We refer to the gamma-ray emission generated by the electrons traveling beyond the core part as the ``wing" components. The steep profile of the halo could be explained by the core component, while the wing components introduce different asymmetric features from the $L_c=\infty$ model. The results show that the expected asymmetry in the $L_c=100$~pc case can be smaller than the $L_c=\infty$ case. Thus, in the absence of significant halo asymmetry found at present\footnote{The recently released H.E.S.S. result also does not indicate significant asymmetry in the Geminga halo \cite{HESS:2023sbf}.}, the possibility of interpreting observations using anisotropic diffusion is enhanced. On the other hand, unlike the $L_c=\infty$ case, the presence of the wing components introduces a certain degree of asymmetry even for $\Phi=0^\circ$. If the anisotropic model is correct, the halo asymmetry may already be detectable using the updated HAWC data \cite{Zhou:2021dgj} or the LHAASO-WCDA data.

In addition, we introduce a semi-analytical method to solve the anisotropic propagation equation, which simplifies anisotropic diffusion to isotropic diffusion through a coordinate transformation. This method is much more convenient than numerical methods.

This work is supported by the National Natural Science Foundation of China under the grants No. 12105292, No. U1738209, and No. U2031110.

\bibliography{references}

\end{document}